# Miniband engineering and topological phase transitions in topological - normal insulator superlattices

G. Krizman[1,2], B.A. Assaf[3], G. Bauer[2], G. Springholz[2], L.A. de Vaulchier[1], Y. Guldner[1]

[1] *Laboratoire de Physique de l'Ecole normale supérieure, ENS, Université PSL, CNRS, Sorbonne Université, 24 rue Lhomond 75005 Paris, France*

[2] *Institut für Halbleiter und Festkörperphysik, Johannes Kepler Universität, Altenberger Strasse 69, 4040 Linz, Austria*

[3] *Department of Physics, University of Notre Dame, Notre Dame, IN 46556, USA*

**Periodic stacking of topologically trivial and non-trivial layers with opposite symmetry of the valence and conduction bands induces topological interface states that, in the strong coupling limit, hybridize both across the topological and normal insulator layers. Using band structure engineering, such superlattices can be effectively realized using the IV-VI lead tin chalcogenides. This leads to emergent minibands with a tunable topology as demonstrated both by theory and experiments. The topological minibands are proven by magneto-optical spectroscopy, revealing Landau level transitions both at the center and edges of the artificial superlattice mini Brillouin zone. Their topological character is identified by the topological phase transitions within the minibands observed as a function of temperature. The critical temperature of this transition as well as the miniband gap and miniband width can be precisely controlled by the layer thicknesses and compositions. This witnesses the generation of a new fully tunable quasi-3D topological state that provides a template for realization of magnetic Weyl semimetals and other strongly interacting topological phases.**

## I. INTRODUCTION

Heterostructures of quantum materials lead to new emergent states of matter beyond what is possible in their bulk form [1–6]. In the case of topological insulators (TIs), theoretical calculations have shown that Dirac, Weyl and nodal line fermions can be artificially created by periodically stacking a TI and a normal insulator (NI) on top of each other [3,7,8]. These superlattices (SL) have been theoretically proposed as novel templates for realization of magnetic Weyl semimetals [3], Weyl superconductors [9], the quantum anomalous Hall phase [8,10–13], flat band superconductivity [5] and strongly interacting topological phases [14]. Experimental realization of TI/NI superlattice structures, however, has posed a formidable challenge. This is mainly due to the limitations in material combinations that are topologically different but well compatible in terms of crystal structures and heteroepitaxial growth [15,16]. As a result, up to now most of the novel phases have been only theoretically predicted, and only recently first experiments have started to explore these effects [17–23].

Here we show that the IV-VI lead tin chalcogenides (Pb,Sn)(Se,Te) topological crystalline insulators (TCIs) [24–26] provide an excellent platform for the realization of artificial TI/NI superlattice structures. This is because their topology [27–30], anisotropy [31,32], band alignment [33,34] and crystal symmetry [35–38] can be controlled on demand by composition, temperature, strain, and/or ferroelectric phase transitions. As a result, band structure engineering of heterostructures can be easily achieved, as has been demonstrated for mid-infrared device applications [39]. In the (Pb,Sn)(Se,Te) TCIs, the non-trivial band topology arises from the band inversion between the $L_6^+$ and $L_6^-$ bands appearing at sufficiently high Sn contents. This leads to the formation of Dirac cone topological surface, respectively, interface states (TIS) that are protected by crystal symmetries [24,25] rather than by time reversal symmetry as in conventional topological insulators [40]. In ultra-thin films and quantum wells, the interface states at the upper and lower film boundaries hybridize, which leads to a gapping of the Dirac cone [27,41,42] as experimentally demonstrated in our previous work [19].

Here, we study TCI/NI superlattice structures with ultra-thin barriers where the topological interface states are not only coupled across the TI quantum wells but also across the normal insulator barrier layers. Using magneto-optical Landau level spectroscopy and envelope function calculations, we demonstrate that due to this coupling, extended topological minibands emerge that can be precisely controlled by growth, temperature, layer thicknesses and compositions. The minibands are directly evidenced by observation of two sets of magnetooptical transitions occurring at the center and edge of the mini Brillouin zone (BZ) imposed by the artificial periodicity of the SL structure. In this way, we reveal that their dispersion, gap size and miniband width can be perfectly controlled by tuning of the coupling constants. The non-trivial miniband topology is experimentally proven by the observation of the topological phase transitions as a function of temperature used as a tuning knob for the band inversion. From our data, we construct for the first time the experimental non-magnetic Burkov-Balents phase diagram [3] predicted for such non-trivial systems. Our results thus represent a text-book topological superlattice system supporting topological minibands artificially designed for various device applications.

## II. GROWTH AND CHARACTERIZATION

Artificial TI/NI superlattice heterostructures were created by molecular beam epitaxy of non-trivial $Pb_{1-x}Sn_xSe$ TI layers (quantum wells) with inverted band gap, alternating with trivial NI $Pb_{1-y-x}Eu_ySn_xSe$ barrier layers. For the topologically non-trivial $Pb_{1-x}Sn_xSe$, $x_{Sn} > 0.21$ was chosen to obtain a negative gap of $2\Delta_{QW} < -20$ meV at 4 K. Alloying of europium into the barrier, on the other hand, turns the band gap *positive* [11], rendering $Pb_{1-y-x}Eu_ySn_xSe$ topologically trivial with a gap $2\Delta_B \sim +150 meV$ for $y_{Eu} \sim 0.05$. Molecular beam epitaxy of $Pb_{1-x}Sn_xSe/Pb_{1-y-x}Eu_ySn_xSe$ superlattices on $BaF_2$ (111) was carried out at a substrate temperature of 360°C under ultra-high vacuum conditions of 5×10⁻¹⁰ mbar using a RIBER 1000 MBE system. PbSe, SnSe, Eu and Se effusion sources were used for growth, and a $Bi_2Se_3$ source for tuning of the carrier concentration to low $10^{18}$ cm$^{-3}$ as determined by Hall effect measurements. The superlattice stacks were grown on a 50 nm $Pb_{1-y}Eu_ySe$ buffer layer pre-deposited on the $BaF_2$ substrate and a 50 nm $Pb_{1-y}Eu_ySe$ on top as a capping layer.

Perfect 2D growth was achieved, evidenced by streak reflection high energy electron diffraction patterns observed throughout superlattice growth. This yields perfect multilayer structures as evidenced by high resolution x-ray diffraction shown in Fig. 1, showing sharp superlattice satellite peaks for all samples. In reciprocal space maps shown in Fig. 1(a,b) theses satellite peaks are perfectly aligned along the $Q_{[111]}$ growth direction, evidencing the very high quality of the samples and full coherency of the interfaces. The resulting superlattice parameters for the investigated samples are listed in Tab. 1.

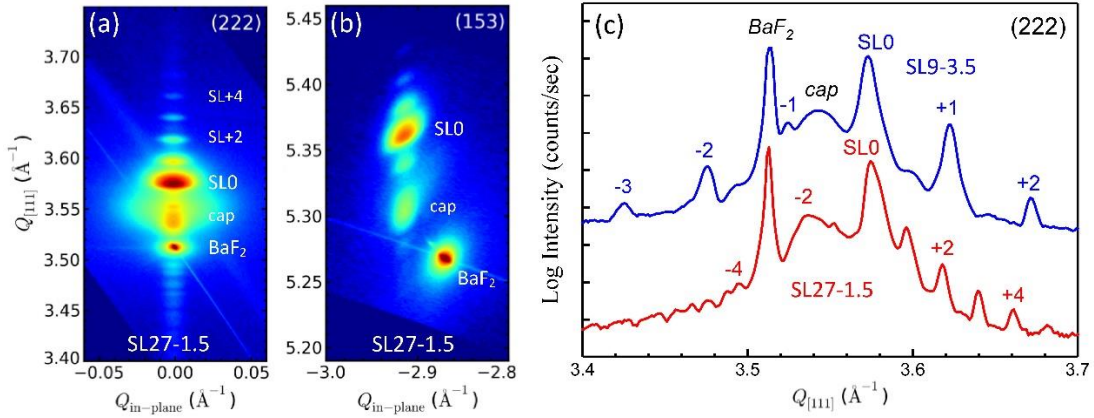

**Figure 1. High resolution x-ray characterization of the TCI/NI superlattice structures**. **(a,b)** Reciprocal space maps of $Pb_{1-x}Sn_xSe$ / $Pb_{1-y-x}Eu_ySn_xSe$ SLs around the symmetric (222) and asymmetric (153) Bragg reflection evidencing perfect pseudomorphic growth. **(c)** Radial diffraction scans along $Q_{[111]}$ normal to the surface for samples SL27-1.5 and SL9-3.5. The satellite peaks are labelled as SLx and the diffraction peaks of the $BaF_2$ substrate and $Pb_{1-y}Eu_ySe$ capping layer are also indicated. The sample parameters obtained are listed in Tab. 1.

### III. MODELING OF THE MINIBANDS

For proper sample design, envelope function theory [44–47] was employed to predict and model the SL band structure. The periodic superlattice potential shown schematically in Fig. 2(a) implies envelope functions satisfying the Bloch theorem with a wave vector $q_z$ that lies within the artificial superlattice BZ reduced within the boundaries $[-\pi/L; +\pi/L]$, determined by the superlattice period $L$. Using a 4-band $\boldsymbol{k}\cdot\boldsymbol{p}$ model, detailed in Appendix A, the topological miniband ($TMB$) dispersions are calculated versus $q_z$ as shown in Fig. 2(b). We find that the electron and hole-like states form mirror-like minibands $TMB(q_z)$ and $TMB'(q_z)$, respectively. Their gap are denoted as $2\delta_0$ and $2\delta_{\pi/L}$ at the center and boundaries of the mini BZ, respectively. Note that due to the multi valley band structure of the IV-VI compounds [34,48–50], the miniband dispersions are slightly different for the oblique and longitudinal valleys (solid and dashed lines in Fig. 2(b)) that are tilted, respectively, aligned parallel to the growth direction. This originates from the admixture of the anisotropic band dispersion of the barriers to the otherwise isotropic $Pb_{1-x}Sn_xSe$ QWs [31,51], which yields slightly different $v_z$, the Dirac velocities along the z//[111] growth direction for the longitudinal and oblique valleys. Note that we define here the Dirac velocity as the slope of the linear part of the $E(k)$ dispersion.

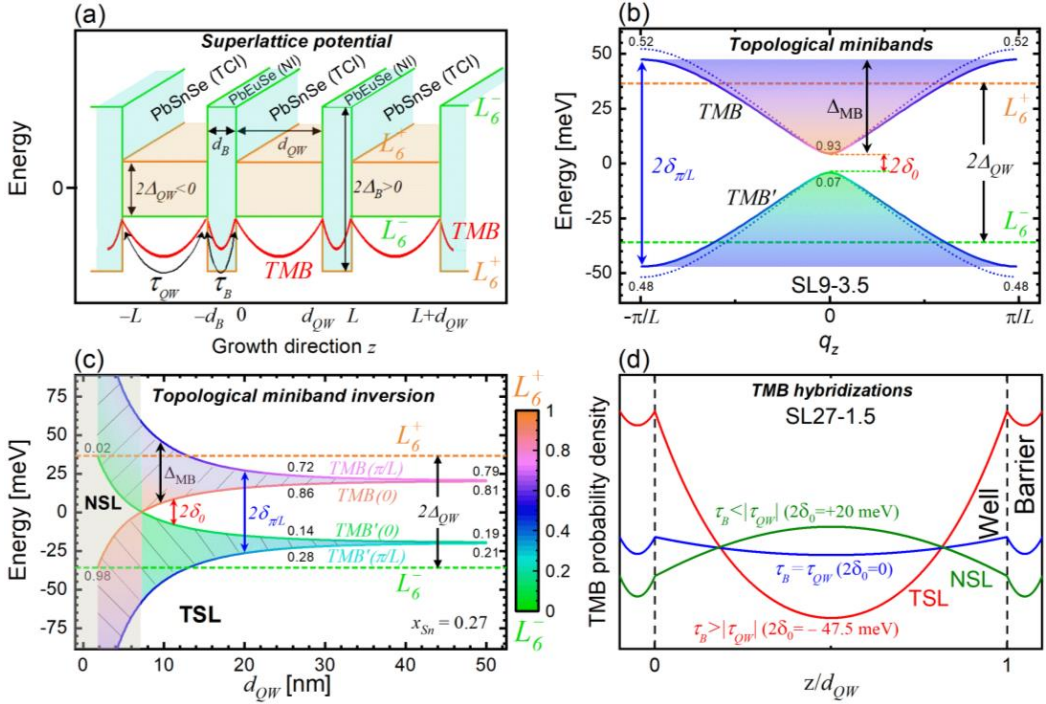

**Figure 2. Topological miniband formation in TCI/NI superlattices. (a)** Modulation of the conduction and valence band edges along the SL structure in the growth direction $z$. The envelope wave function of the topological miniband concentrated at the interface is illustrated by the red curve, the black arrows indicate the intrawell and interwell tunnel coupling $\tau_{QW}$ and $\tau_B$, respectively. **(b)** Miniband dispersion $E(q_z)$ for the longitudinal (solid lines) and oblique valleys (dashed lines) derived by $\boldsymbol{k}\cdot\boldsymbol{p}$ theory at $k_x = k_y = 0$ and 4.2 K for the superlattice structure SL9-3.5 listed in Tab. 1. The color scale represents the symmetry ($L_6^+$ versus $L_6^-$) of the minibands (see scale bar in (c)). Label numbers denote the $L_6^+$ proportion. **(c)** Evolution of the minibands and their symmetry (color scale) in the conduction and valence band (($TMB$, respectively, $TMB'$) as a function of Pb$_{1-x}$Sn$_x$Se thickness $d_{QW}$. The barrier width is fixed to $d_B = 3.5$ nm and the Pb$_{1-x}$Sn$_x$Se composition to $x_{Sn} = 0.27$. The corresponding bulk band gaps are $2\Delta_{QW} = -72.5$ meV (dashed horizontal lines) and $2\Delta_B = +150$ meV. **(d)** Probability density of the topological miniband envelope wave function across the SL structure for different miniband topologies. Green line: Normal superlattice (NSL) with $2\delta_0 = +20$ meV and $\tau_B < |\tau_{QW}|$, blue line: Zero gap SL with $2\delta_0 = 0$ and $\tau_B = |\tau_{QW}|$, red line: Topological superlattice (TSL) with $2\delta_0 = -47.5$ meV and $\tau_B > |\tau_{QW}|$. The different character is set by changing the Pb$_{1-x}$Sn$_x$Se band gap from $2\Delta_{QW} = 10, -10, -60$ meV.

Solution of the $\boldsymbol{k}\cdot\boldsymbol{p}$ Hamiltonian yields the size of the miniband gaps $2\delta_0$ and $2\delta_{\pi/L}$ as a function of layer thicknesses and compositions. The results are exemplified in Fig. 2(c), where the evolution of the minibands is shown as a function of QW thickness for a fixed barrier thickness $d_B = 3.5\ nm$ and bulk band gaps set to $2\Delta_{QW} = -72.5$ meV and $2\Delta_B = +150$ meV, respectively. Evidently, the gap between the minibands goes to zero at a critical QW thickness (vertical line), indicating that only at sufficiently large $d_{QW}$ the symmetry of the band edges is inverted.

Based on the solution of the $\boldsymbol{k}\cdot\boldsymbol{p}$ model, we find that in a good approximation, the $\delta(q_z)$ dispersion of the miniband edges is given by (see Appendix B):

$$\delta(q_z) \cong \sqrt{\frac{2\hbar^2 v_z^2 [1 - \cos(q_z L)]}{L^2} + \delta_0^2} \quad (1a)$$

$$\delta_0 = \delta(q_z = 0) \cong \frac{d_{QW}\Delta_{QW} + d_B\Delta_B}{L} \quad (1b)$$

This means that the hybridization gap $2\delta_0$ of the superlattice structure is essentially equal to the average of QW and barrier band gaps weighted according to their layer thickness. This is due to the close similarity of the band parameters of the layers. According to Eq. (1), the minibands can be fully designed by the superlattice structure. Most importantly, the gap assumes a *negative* value only under the condition that $d_B\Delta_B < |d_{QW}\Delta_{QW}|$ and $\Delta_{QW}$ is negative. This means that a non-trivial miniband topology is not simply formed when the band gap of the TI well material is inverted.

We further identify the *intra*well and *inter*well coupling strength $\tau_{QW}$ and $\tau_B$ between the TI/NI interfaces as $\tau_{QW} \equiv -\hbar v_z/d_{QW}\Delta_{QW}$ and $\tau_B \equiv \hbar v_z/d_B\Delta_B$, respectively. Note that in Dirac matter, the penetration depth of the topological state is given by $\lambda_{QW} = \hbar v_z/|\Delta_{QW}|$ or $\lambda_B = \hbar v_z/\Delta_B$ in the well and in the barrier respectively [19,52,53]. This means that $\tau_{QW} = \lambda_{QW}/d_{QW}$ and $\tau_B = \lambda_B/d_B$ capture the recovery of the topological state wavefunctions coming from two distinct interfaces separated by $d_{QW}$ or $d_B$. We can then define the strong coupling limit when $|\tau_{QW,B}| > 1$, which insures significant interactions between interface states.

Using the expression of $\tau_{QW}$ and $\tau_B$, Eq. (1b) simplifies to $\delta_0 = \hbar v_z(\tau_B^{-1} - \tau_{QW}^{-1})/L$. The topological phase transition where $\delta_0$ is zero then occurs under the condition that:
$$d_{QW}\Delta_{QW} + d_B\Delta_B = 0 \quad \text{or} \quad \tau_B = \tau_{QW} \quad (2)$$

This leads to a change of the symmetry of the miniband states ($L_6^+$ versus $L_6^-$) that is represented by the color scale in Fig. 2(c) as further detailed in the Appendix C. This shows that indeed above a critical thickness where $\tau_B > |\tau_{QW}|$, the character of the minibands is inverted, i.e., the SLs become topologically non-trivial [3,7]. The topological character is thus encoded in the $|\tau_{QW}/\tau_B|$ ratio that is $< 1$ for the nontrivial, but $> 1$ for trivial structures.

The pronounced effect of the band topology on the wave functions of the minibands is demonstrated by Fig. 2(d), where the wave function probability density across the QW and barriers is depicted for three cases, i.e., a normal SL (NSL) with $2\delta_0 > 0$, a zero gap SL and a topological superlattice (TSL) with $2\delta_0 < 0$. Whereas, for the NSL with $\tau_B < |\tau_{QW}|$ (green line in Fig. 2(d)), the probability density is maximal in the center of the QW as it is generic for conventional semiconductor superlattices, this is exactly opposite for the TSL with $\tau_B > |\tau_{QW}|$ (red line), where the probability density is minimal in the QW and high in the barriers. This remarkable difference is a clear signature of the topological character of the structure. At the phase transition between the NSL and TSL cases, the miniband gap is zero and $\tau_B = |\tau_{QW}|$. As a result, the average probability density is the same in the QWs and barriers (blue line in Fig. 2(d)), i.e., the electrons/holes are evenly distributed over the QWs and barriers and thus most delocalized in the SL structure.

The coupling strengths $\tau_{QW}$ and $\tau_B$ directly control the miniband widths $\Delta_{MB}$ (shaded regions in Fig. 2(c)). Accordingly, when layer thicknesses are reduced to few nanometers, a strong coupling regime emerges witnessed by a drastic widening of the minibands. Conversely, when the wells and/or barriers are thick, the coupling between the interfaces is diminished, narrowing the minibands as seen in Fig. 2(c). In the thick barrier limit, uncoupled quantum wells are formed in which the interface states do no longer hybridize across the barrier layers [19]. Accordingly, not only the band gaps but also miniband widths can be engineered by control of the layer thicknesses.

For our magneto-optical experiments, two types of samples were prepared, namely, TSL designated to exhibit a *negative* miniband gap and nontrivial topology (samples SL9-3.5 and SL27-1.5, see Tab. 1), and NSL with reduced Sn content. The Sn content is a crucial parameter that controls the well potential depth. It is reduced for SL15-3.5, rendering the SL gap *positive*. SL15-3.5 is therefore a control sample with trivial topology. In the samples, the QW and barrier thicknesses were varied to yield different coupling strengths and miniband widths. The miniband gaps $2\delta_0$ and coupling ratios $|\tau_{QW}/\tau_B|$ at 4 K are also listed in Tab. 1, where $|\tau_{QW}/\tau_B| < 1$ for topological SL9-3.5 and SL27-1.5, but $> 1$ for the trivial SL15-3.5 reference sample. It is noted that due to the small lattice mismatch between the QW and barrier materials, a small tensile strain is imposed in the QWs and the barriers are slightly compressed. This leads to small deviations in the band gaps $\Delta_{QW}$ and $\Delta_B$ with respect to the unstrained bulk material [31,34].

**Table 1.** Sample parameters of the investigated TCI/NI superlattice structures, composed of $Pb_{1-x}Sn_xSe$ TCI QWs alternating with trivial NI $Pb_{1-y-x}Eu_ySn_xSe$ barriers, repeated $N$ times. Also listed are the effective miniband gaps $2\delta_0$ and coupling ratios $|\tau_{QW}/\tau_B|$ (Eq. (1)) for $T$ = 4.2 K. $2\delta_0 < 0$ and $|\tau_{QW}/\tau_B| < 1$ indicate non-trivial topology, whereas for $2\delta_0 > 0$ and $|\tau_{QW}/\tau_B| > 1$, the superlattices are topologically trivial. The well and barrier material band gaps $2\Delta_{QW}$ and $2\Delta_B$ are obtained from the fits of the magneto-optical data, which are shown later.

| Parameter | SL9-3.5 | SL27-1.5 | SL15-3.5 |
|---|---|---|---|
| TCI: $d_{QW}$ [nm] | $9 \pm 0.2$ | $27 \pm 0.2$ | $15 \pm 0.2$ |
| $x_{Sn}$ in $Pb_{1-x}Sn_xSe$ | $0.27 \pm 0.01$ | $0.26 \pm 0.01$ | $0.22 \pm 0.01$ |
| Band gap $2\Delta_{QW}$ at 4 K [meV] | $-72.5$ | $-60$ | $-20$ |
| NI: $d_B$ [nm] | $3.5 \pm 0.2$ | $1.5 \pm 0.2$ | $3.5 \pm 0.2$ |
| $y_{Eu}$ in $Pb_{1-y-x}Eu_ySn_xSe$ | 0.05 | 0.05 | 0.05 |
| Band gap $2\Delta_B$ at 4 K [meV] | 150 | 140 | 150 |
| SL: Number $N$ of periods | 40 | 20 | 30 |
| Miniband gap $2\delta_0$ at 4 K [meV] | -10 | -47.5 | +10 |
| $|\tau_{QW}/\tau_B|$ at 4 K | 0.80 | 0.14 | 1.75 |

### IV. LANDAU LEVEL SPECTROSCOPY OF TOPOLOGICAL MINIBANDS

To assess the topological minibands magneto-optical spectroscopy was performed in Faraday geometry at magnetic fields up to 15 T [19,43]. The results for SL9-3.5 and SL15-3.5 are shown in Fig. 3 for $T$ = 4.2 and 160 K. In both cases, a large number of Landau level transitions are observed (arrows in Fig. 3(a,b)), shifting to higher energies as the magnetic field increases. From this we construct fan charts shown in Fig. 3(c,d), where each data point corresponds to a minimum in the transmission spectra in (a,b). The fan charts are analyzed using Landau level transitions obtained from $\boldsymbol{k}\cdot\boldsymbol{p}$ calculations presented in the Appendix D (solid and dashed lines) and fitted to the experimental data. From the analysis, we unambiguously identify the existence of *two* individual subsets of transitions, indicated in Fig. 3 by the red and blue colors. These are identified to occur at the miniband extrema at $q_z = 0$ (red) and $q_z = \pm\pi/L$ (blue), respectively, where the joint density of states of the minibands is maximal. Each extrapolates to a different energy at $B = 0$, corresponding to the miniband gaps $2\delta_0$ and $2\delta_{\pi/L}$ at the center and edge of the superlattice BZ respectively – in perfect agreement with the $\boldsymbol{k}\cdot\boldsymbol{p}$ calculations. The validity of the assignment is evidenced by the perfect fit obtained in each case, with the fit parameters listed in the Supplementary Material [54]. Most importantly, the observation of the two independent series of transitions directly proves the miniband formation in our strongly coupled superlattice structures.

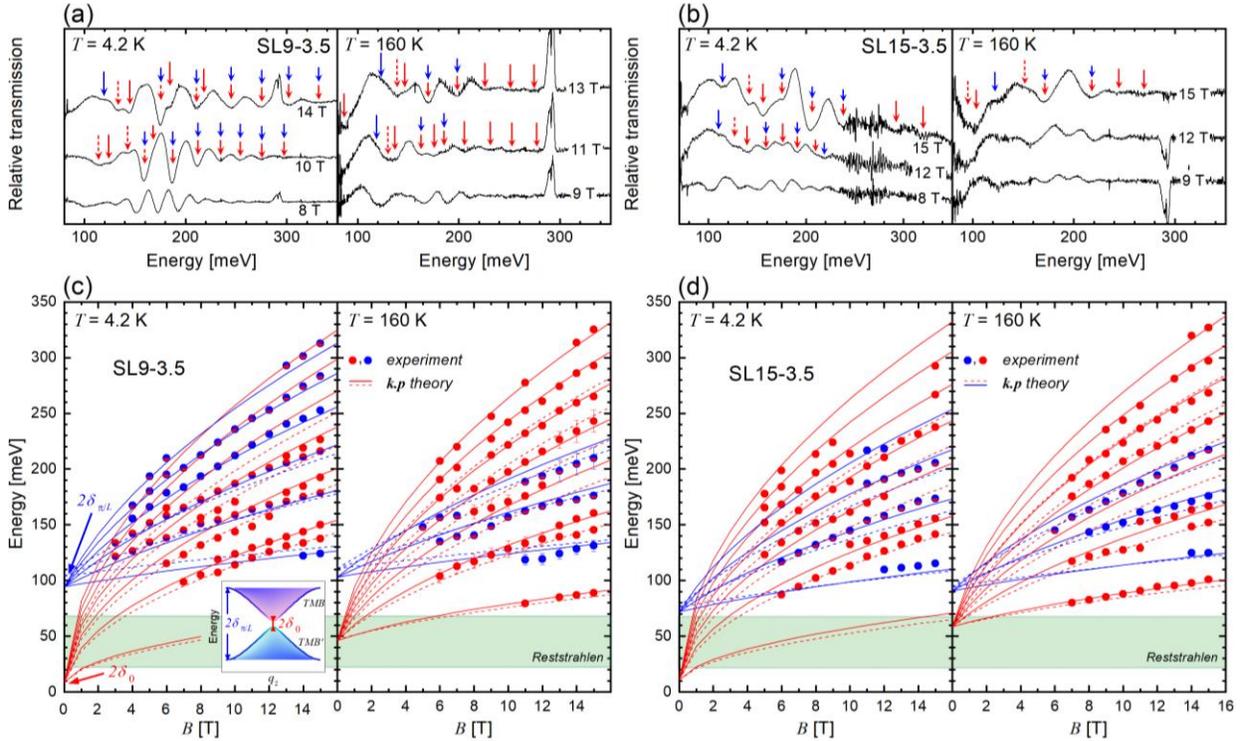

**Figure 3. Magnetooptical spectroscopy. (a,b)** Normalized transmission spectra of the superlattice samples SL9-3.5 and SL15-3.5 at 4.2 and 160 K at different magnetic fields of up to $B = 15$ T. The minima are due to Landau level transitions between the minibands at $q_z = 0$ and $q_z = \pi/L$, marked by red and blue arrows, respectively. **(c,d)** Magnetooptical fan charts derived from the experiments (red/blue dots) compared to the calculations by the $\boldsymbol{k}\cdot\boldsymbol{p}$ model for the longitudinal and oblique valleys (solid and dashed lines, respectively). The extrapolated transition energies $2\delta_0$ and $2\delta_{\pi/L}$ at $B=0$ of the two transition sets are indicated by the arrows and in the insert. The green shaded regions indicate the experimentally non-accessible energy range blocked by the reststrahlen band of the substrate and window cut-offs.

For the superlattice SL9-3.5, the extrapolation of the data points yields a miniband gap of $|2\delta_0| = 10 \pm 5$ meV at 4.2 K, whereas the second set of transitions yields a gap of $|2\delta_{\pi/L}| = 95 \pm 5$ meV at the boundary of the BZ. Both values perfectly agree with the calculated values in Fig. 2(c) for the given TI/NI layer thicknesses. Using $\Delta_{MB} = |\delta_0 - \delta_{\pi/L}|$, a miniband width of $42.5 \pm 10$ meV is derived for this sample. According to the $\boldsymbol{k}\cdot\boldsymbol{p}$ calculations (cf. Fig. 2(c)), the miniband gap is inverted, i.e., the minibands are topologically non-trivial at 4.2 K. This is supported by the fact that the in-plane Dirac velocities, determined from the fits for the longitudinal and oblique valleys ($v_\parallel^l = 4.40 \times 10^5$ m/s and $v_\parallel^o = 4.10 \times 10^5$ m/s) are below the critical values where the band gap is inverted [32]. It is noted that the small difference in the in-plane miniband dispersion for the longitudinal and oblique valleys arises from the admixture of the band anisotropy of the $Pb_{1-y-x}Eu_ySn_xSe$ barriers to that of the QWs caused the large extent of the wave function across the SL period, as the bulk bands of $Pb_{1-x}Sn_xSe$ with $0.21 < x_{Sn} < 0.29$ are otherwise isotropic. This is particularly pronounced for topological superlattices because the probability density of the wave function is strongly enhanced within the barriers as shown by Fig. 2(d). For the second SL sample SL15-3.5 with lower Sn content, the same analysis yields $2\delta_0 = +10 \pm 5$ meV and $2\delta_{\pi/L} = 74 \pm 5$ meV, rendering the gap positive and the minibands topologically trivial. Moreover, due to the increased QW thickness of $d_{QW} = 15$ nm, the miniband width is reduced to $\Delta_{MB} = 32 \pm 10$ meV, following nicely the trend of Fig. 2(c).

We highlight that the observed magnetooptical transitions occur both at $q_z = 0$ and $q_z = \pi/L$ where the joint density of states for optical transitions is largest. The fact that both transitions are simultaneously observed and precisely fit to the calculations evidences the existence of minibands in the superlattice structures. To this end, we emphasize that the higher energy transitions (marked in blue) cannot be interpreted as transitions between higher energy excited states or higher order minibands. Indeed, for the samples in Fig. 3, these would lie in the continuum above the band gap $2\Delta_B$ of the barrier material, as shown in the supplementary material. Consequently, we safely attribute the observed absorptions to the emergent minibands caused by the hybridization of interface states. To the best of our knowledge, this type of artificial energy bands TI/NI multilayer structures has not been realized before and in any topological material system.

## V.  EXPERIMENTAL DEMONSTRATION OF SYMMETRY INVERSION

The topological character of the superlattice structures is directly revealed by magnetooptical measurements at varying temperatures in which the fundamental band gaps of the QW and barrier materials are tuned [27,31,32]. First, we focus on the superlattice SL27-1.5 that displays the largest negative miniband gap at 4.2 K and is thus, most deeply in the inverted topological region. The magnetooptical fan chart of the sample shown in Fig. 4(a) yields a miniband gap of $2\delta_0 = -47.5 \pm 2.5$ meV and $2\delta_{\pi/L} = -75 \pm 5$ meV at 4.2 K as indicated by the solid lines extrapolated to $B = 0$. Figure 4(b) displays the corresponding transmission spectra at a fixed magnetic field of 15 T as a function of temperature from 4.2 to 225 K. Clearly, the first interband transition occurring at $q_z = 0$, highlighted by the red dots, exhibits a non-monotonic behavior in its position, shifting initially to lower energies as the temperature increases, but then reverses and shifts in the opposite direction above 160 K.

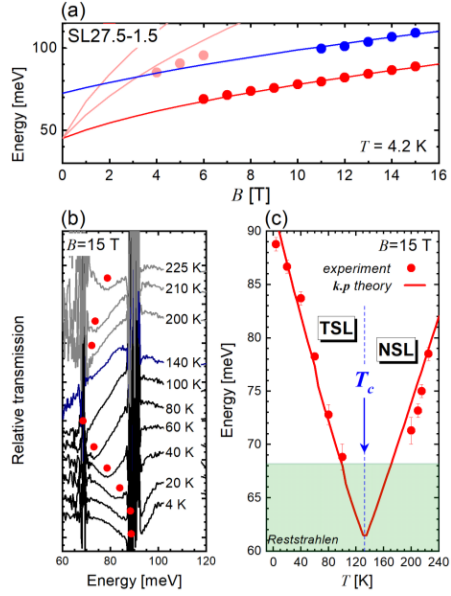

**Figure 4. Temperature dependence of the miniband gap. (a)** Magnetooptical fan chart of superlattice SL27-1.5 at 4.2 K, showing the ground transitions between the minibands at $q_z = 0$ (red) and $|q_z| = \pi/L$ (blue) on an enlarged scale. **(b)** Temperature dependence of the far-infrared transmission spectra at $B = 15$ T in which the lowest energy transition is indicated by the red dots. The energy position of this transition is shown in **(c)** as a function of temperature together with the theoretical fit (solid line) obtained by the $\mathbf{k.p}$ model. The critical temperature $T_c \cong 130$ K, indicated by the blue arrow, separates the TSL from the NSL phases and where $\partial|\delta_0|/\partial T$ changes sign. Note that transitions below 70 meV are masked by the reststrahlen band of the substrate.

The shifts are summarized in Fig. 4(c), where the experimental data (red dots) is compared with the $\mathbf{k.p}$ calculations (red line). Evidently, the non-monotonic shift is perfectly reproduced by our model. The effect originates from the anomalous temperature dependence of the band gaps of IV-VI materials in which the TCI band inversion is induced by the $sp$ repulsion between the $L$ bands and lower lying $S$ bands rather than by spin-orbit coupling [55]. This repulsion decreases with increasing interatomic distances, which lifts the band inversion and renders the material trivial as the temperature is increased. The same also occurs in our TCI/NI structures, albeit at a different critical temperature $T_c$ because the superlattice miniband inversion is not only governed by the bulk bands, but also by the thicknesses of the well and barrier materials (Eq. (1)). Most importantly, the abrupt sign change of $\partial|\delta|/\partial T$ of the TSL structure is a clear evidence for the occurrence of this topological phase transition, with a critical temperature $T_c \approx 130$ K in this sample, below which $\partial|\delta|/\partial T$ is negative. This is the hallmark for the topological nature of our TCI/NI superlattice system.

The non-monotonic behavior is to be contrasted with our previous observations for TCI multi quantum well structures [19], where due to an order of magnitude wider barriers ($d_B > 35$ nm) the coupling between the topological interface states across the barriers is negligible ($\tau_B \approx 0$). As a result, no minibands are formed and the hybridization gap of TIS states only monotonically increases with temperature and thus, no sign changes occurs. In contrast, for the presently studied strong coupled structures, our experiments reveal that extended minibands are formed due to the strong interwell hybridization $\tau_B > 0$, inducing an emergent topological phase transition.

The complete data set for all superlattice samples is summarized in Fig. 5(a-c), which shows the evolution of the miniband gaps $2\delta_0$ and $2\delta_{\pi/L}$ (red and blue dots) as a function of temperature together with the $\mathbf{k}\cdot\mathbf{p}$ calculations (solid lines). For all cases, theory and experiments perfectly fit to one another. The bulk band gaps $\Delta_{QW}(x,T)$ and $\Delta_B(y,T)$ obtained from the magnetooptical fits are shown as well for comparison by the open circles in Fig. 5(a-c), and they agree very well with the empirical expressions (dashed lines) derived in our previous works [31,34]. For the two superlattices SL9-3.5 and SL27-1.5, the minibands are inverted at cryogenic temperatures and thus, they are topological non-trivial. Their absolute miniband gap value of $|2\delta_0|$ decreases with temperature and turns to positive values when the critical temperature $T_c$, indicated by the vertical dashed lines, is reached. This marks the topological phase transition from a TSL to an NSL structure.

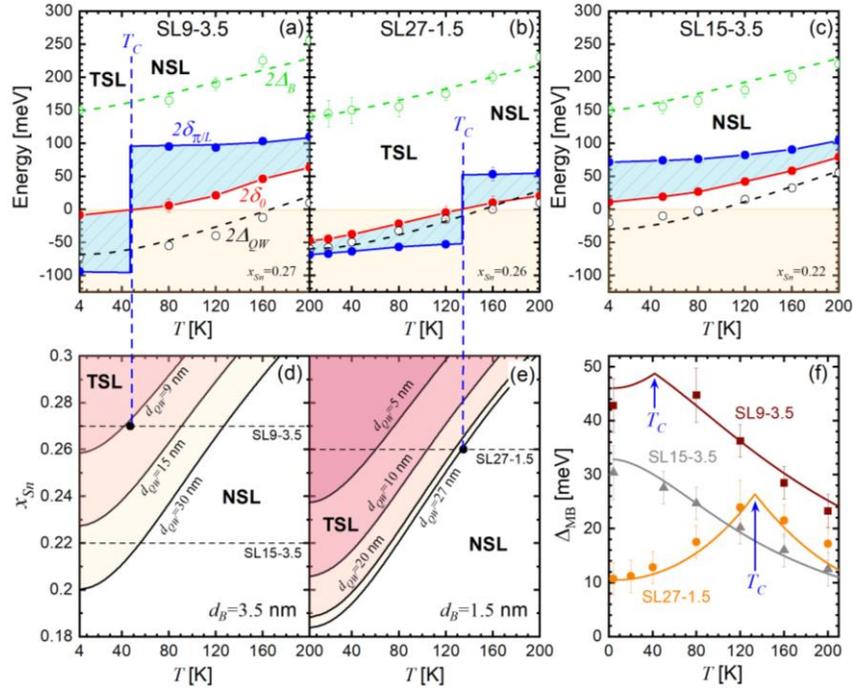

**Figure 5. Demonstration of topological phase transitions. (a-c)** Temperature dependence of the superlattice miniband gaps $2\delta_0$ (red) and $2\delta_{\pi/L}$ (blue) at $q_z = 0$ and $\pm\pi/L$ obtained by experiments (dots) and $\mathbf{k}\cdot\mathbf{p}$ model (solid lines). The region of the miniband band inversion is highlighted in yellow. The vertical dashed lines indicate the critical temperature $T_c$ below which the SLs are nontrivial. Above $T_c$ they are trivial. This transition is due to a symmetry inversion which changes the sign of the miniband gaps. The blue shaded area represents twice the miniband width $2\Delta_{MB}$. Also shown is the temperature dependence of the band gaps of the QW ($2\Delta_{QW}(x,T)$, black circles) and the barriers ($2\Delta_B(y,T)$, green circles) obtained from the fits that nicely agree with our previous work (dashed lines) [31,34]. **(d,e)** Topological phase diagrams of the SL structures as a function of temperature and Pb$_{1-x}$Sn$_x$Se composition for fixed barrier thickness of $d_B = 3.5$ nm (d) and 1.5 nm (e). The solid lines represent the phase boundaries for different QW thicknesses $d_{QW}$ and the shaded regions indicate TSL phases. The black dots in the phase diagrams mark the experimental phase transitions observed for our samples. **(f)** Temperature dependence of the miniband width $\Delta_{MB}$ derived from experiments (symbols) and the $\mathbf{k}\cdot\mathbf{p}$ model (solid lines). The cusps mark the topological-to-normal insulator superlattice phase transition as indicated by the arrows.

The derived $T_c$ values are 40 K and 130 K, respectively, which nicely agrees with the $\boldsymbol{k}\cdot\boldsymbol{p}$ calculations (solid lines). The difference in $T_c$ is mainly due to the different barrier thicknesses (see Tab. 1), for which reason the ratio $|\tau_{QW}/\tau_B|$ of SL9-3.5 is closer to 1 at 4.2 K than the one of SL27-1.5. The resulting weaker interwell coupling in SL9-3.5 makes it closer to the topological phase transition. For the third sample SL15-3.5 (Fig. 5(c)), the topological phase transition does not occur because it remains in the NSL phase down to 4.2 K, where $2\delta_0 = +10$ meV is still positive. Thus, it serves as a control sample that unequivocally reveals that the topological nature of the TSL is intrinsically coupled with the temperature dependent topological phase transition. Moreover, it shows that the SL structures can be topologically trivial even if the quantum wells are in the topological crystalline insulator state.

Finally, we want to highlight that the topological character of the SL system is also encoded in the miniband widths $\Delta_{MB}$ represented by the shaded regions in Fig. 5(a-c). For the TSLs in (a,b), the miniband width increases with increasing temperature, displaying a maximum at the TSL/NSL transition (arrows Fig. 5(a,b)) and thereafter again decreasing, whereas for the NSL (SL15-3.5, Fig. 5(c)) the minibands width only monotonically decreases. Accordingly, for the TSL, $\Delta_{MB}(T)$ displays a cusp at the topological phase transition as shown in Fig. 5(f), both in theory (solid lines) and experiments (dots). This observation is therefore another clear-cut criterion for a topological phase transition in SLs. The cusp arises from the fact that the intra- and interwell coupling strengths are equal when the gap $2\delta_0 = 0$ and thus, the miniband wave functions are maximally delocalized, maximizing the miniband width. To this end, we refer to Fig. 2(d), which illustrates the calculated probability density of SL27-1.5 $TMB$ at 4.2 K (in red), $T_C = 130$ K (in blue) and 200 K (olive). In fact, at the topological phase transition where $2\delta_0 = 0$, the miniband width scales as $\Delta_{MB} \cong 2\hbar v_z/L$ following Eq. (1a). Remarkably, it is essentially independent of the QW and barrier thicknesses.

The topological phase transitions are put into broader perspective by the topological phase diagrams of Fig. 5(d,e). These display the topological state and the phase boundaries between the TSL and NSL structures as a function of QW composition and temperature for different QW thicknesses (solid lines) but fixed barrier thickness $d_B = 3.5$ and 1.5 nm. As indicated by the experimental data points (black dots) obtained from Fig. 5(a,b), these phase diagrams are in excellent agreement with our experiments. Therefore, they accurately describe the topological character of the system and serve as guides for engineering the miniband properties for a given application.

## VI. CONCLUSION

Using Landau level spectroscopy, we have demonstrated the formation of topological minibands in artificial TCI/NI superlattices obtained by molecular beam epitaxy and band structure engineering of IV-VI semiconductor heterostructures. By envelope function calculations we revealed that the minibands are the offspring of the hybridized topological interface states that tunnel couple both across the normal insulator barrier layers as well as across the TI quantum wells. In the topological SL state, this gives rise to a pronounced shift of the wave function envelope from the quantum wells to the barriers, which discriminates the TSL from NSL structures. As a result, the topological phase as well as miniband gap dispersions can be perfectly controlled by the layer thicknesses and compositions, and tunable miniband gaps and miniband widths are attained. The temperature-induced phase transition of the miniband topological character is in perfect agreement with our theoretical model. Thereby, we experimentally demonstrate for the first time the recently predicted Burkov-Balents phase diagram [3]. Accordingly, our TI/NI superlattices provide a new quasi-3D topological state that can be engineered over a wide range, which offers new avenues towards non-zero dissipation less spin Hall currents [3]. Moreover, by breaking time reversal symmetry using magnetic doping [38,56–58], magnetic topological superlattices with tunable Weyl, or even line node semimetal phases [3,7,8] can be reached.

## ACKNOWLEDGEMENTS

The authors sincerely thank G. Bastard and R. Ferreira for fruitful discussions and the ANR N° ANR-19-CE30-022-01 and Austrian Science Fund FWF, Project I-4493 for financial support. B. A. A. is partly supported by NSF-DMR-1905277.

## APPENDIX A: $k.p$ MODEL FOR SUPERLATTICES

The stacking of Pb$_{1-x}$Sn$_x$Se and Pb$_{1-y-x}$Sn$_x$Eu$_y$Se layers makes $z$-dependent $L_6^-$ and $L_6^+$ band edges, where $z$ is the growth axis. At the interfaces between two layers, the bands of same symmetry must be connected [44] thus, leading to a potential that inverses the conduction and valence band at each interface. As the system is considered electron-hole symmetric [34], this potential can be modelled by a $z$-dependent energy gap $\Delta(z)$ that changes sign across each interface. Due to confinement, $k_z$ is not a good quantum number and is replaced by its operator value $-i\,\partial/\partial z$. The $k^2$-terms coming from the interactions between $L_6^-$ or $L_6^+$ with other bands located at much higher or lower energies are neglected [59]. We also consider a magnetic field along the $z$-axis. In this way, for $n > 0$ ($n$ being the Landau level index) the Hamiltonian of the SL system can be written in the basis $L_6^+\alpha\,|n-1\rangle$; $L_6^+\beta|n\rangle$; $L_6^-\alpha|n-1\rangle$; $L_6^-\beta|n\rangle$ ($|n\rangle$ being the harmonic oscillator functions) as [19,48,60]:

$$\begin{pmatrix} -\Delta(z) & 0 & -i\hbar v_z \dfrac{\partial}{\partial z} & v_\parallel \sqrt{2e\hbar Bn} \\ 0 & -\Delta(z) & v_\parallel \sqrt{2e\hbar Bn} & i\hbar v_z \dfrac{\partial}{\partial z} \\ -i\hbar v_z \dfrac{\partial}{\partial z} & v_\parallel \sqrt{2e\hbar Bn} & \Delta(z) & 0 \\ v_\parallel \sqrt{2e\hbar Bn} & i\hbar v_z \dfrac{\partial}{\partial z} & 0 & \Delta(z) \end{pmatrix} \quad (A1)$$

where $\alpha$ and $\beta$ are the spins with underlying spin-orbit coupling; $v_z$ and $v_\parallel$ are the electron velocities respectively along and perpendicular to the growth direction. The $j$-th energy and wavefunctions of the confined states at $k_x = k_y = 0$ are calculated by reducing the Hamiltonian (A1) accordingly. This yields two spin-decoupled equations:

$$\begin{pmatrix} -\Delta(z) - E_j & \xi i\hbar v_z \dfrac{\partial}{\partial z} \\ \xi i\hbar v_z \dfrac{\partial}{\partial z} & \Delta(z) - E_j \end{pmatrix} \begin{pmatrix} F_1^{(j)} \\ \xi F_2^{(j)} \end{pmatrix} = 0$$

where $\xi = \pm$ represents the spins and $E_j$ denotes the spin-degenerated energy of the $j$-th confined states and goes along with their two-component spinor envelope wavefunctions. The envelope wavefunction of the $j$-th confined states have a $L_6^+$ and a $L_6^-$ component: $F_1^{(j)}$ and $F_2^{(j)}$ respectively.

In order to calculate each component of each envelope function, the current probability continuity conditions are applied at each interface for the $L_6^+$ component $F_1^{(j)}$. Therefore, at the interface between the well and the barrier, $F_1^{(j)}$ must be continuous as well as the quantity [45,46]:

$$\frac{1}{\Delta(z) - E_j} \frac{\partial F_1^{(j)}}{\partial z} \tag{A2}$$

$F_2^{(j)}$ is then deduced from $F_1^{(j)}$ by:

$$F_2^{(j)}(z) = \frac{i\hbar v_z}{\Delta(z) - E_j} \frac{\partial F_1^{(j)}}{\partial z}$$

The SL periodicity implies that $F_1^{(j)}$ can be written as a Bloch wave: $F_1^{(j)}(z+L) = F_1^{(j)}(z)e^{iq_z L}$ with $-\pi/L < q_z < +\pi/L$. If $|E_j| < |\Delta_{QW}|$, the continuity of $F_1^{(j)}$ and (A2) at $z = d_{QW}$ and $z = L$ yields a four-equation system giving the following secular equation:

$$\cos(q_z L) = \cosh(\kappa d_{QW})\cosh(\rho d_B) - \frac{1}{2}\left(\gamma + \frac{1}{\gamma}\right)\sinh(\kappa d_{QW})\sinh(\rho d_B) \tag{A3}$$

with $\gamma = -\frac{\kappa}{\rho}\frac{E_j - \Delta_B}{E_j - \Delta_{QW}}$; $\kappa = \frac{1}{\hbar v_z}\sqrt{\Delta_{QW}^2 - E_j^2}$; and $\rho = \frac{1}{\hbar v_z}\sqrt{\Delta_B^2 - E_j^2}$.

Here, we have written $F_1^{(j)}$ as:

$$F_1^{(j)}(z) = \begin{cases} a\cosh(\kappa z) + b\cosh(\kappa z), & \text{in the well} \\ c\cosh(\rho(z - d_{QW})) + d\cosh(\rho(z - d_{QW})), & \text{in the barrier} \end{cases}$$

where $a, b, c, d$ are the four eigenvector components of the system. For a negative enough $\Delta_{QW}$, (A3) gives two solutions that describe $TMB$ and $TMB'$ dispersions versus $q_z$ at $k_x = k_y = 0$, as it is shown in Fig. 2(b,c). $F_1^{(j)}$ is then deduced and shown for instance in Fig. 2(d). For miniband with $|E_j| > |\Delta_{QW}|$, (A3) is transformed into:

$$\cos(q_z L) = \cos(k d_{QW})\cosh(\rho d_B) - \frac{1}{2}\left(\tilde{\gamma} - \frac{1}{\tilde{\gamma}}\right)\sin(k d_{QW})\sinh(\rho d_B)$$

with $\tilde{\gamma} = \frac{k}{\rho}\frac{E_j - \Delta_B}{E_j - \Delta_{QW}}$ and $k = \frac{1}{\hbar v_z}\sqrt{E_j^2 - \Delta_{QW}^2}$.

## APPENDIX B: APPROXIMATION OF THE MINIBAND DISPERSION

Equation (A3) can be approximated if the cosh and sinh functions are developed to the 2<sup>nd</sup> order. One gets with $E_j = \delta$:

$$\cos(q_z L) \cong 1 + \frac{\kappa^2 d_{QW}^2}{2} + \frac{\rho^2 d_B^2}{2} - \frac{1}{2}\left(\gamma + \frac{1}{\gamma}\right)\kappa d_{QW} \rho d_B$$

$$\Leftrightarrow \delta(q_z) \cong \sqrt{\frac{2\hbar^2 v_z^2[1 - \cos(q_z L)] + (d_{QW}\Delta_{QW} + d_B\Delta_B)^2}{L^2}}$$

which is Eq. (1a,b). The approximation $\rho d_B \sim 0$ is justified as the investigated SL have ultrathin barriers. Small values of $\kappa d_{QW}$ are obtained if the wells are thin or if $\delta$ is close to $\Delta_{QW}$, which is the case in the present work.

## APPENDIX C: SYMMETRY INVERSION OF THE MINIBANDS

A symmetry inversion can be induced by the interwell coupling $\tau_B$ in a superlattice. Indeed, one can notice that the SL gap $2\delta_0$ is vanishing at a certain point (see Fig. 2(c) for instance). Having a bound state at the middle of the quantum well energy gap implies that $E_j = 0$ and therefore $\kappa = \frac{|\Delta_{QW}|}{\hbar v_z}$, $\rho = \frac{\Delta_B}{\hbar v_z}$ and $\gamma = 1$. Equation (A3) thus becomes:

$$\cos(q_z L) = \cosh(\kappa d_{QW})\cosh(\rho d_B) - \sinh(\kappa d_{QW})\sinh(\rho d_B) = \cosh(\kappa d_{QW} - \rho d_B)$$

We conclude that for $d_B \Delta_B = d_{QW}|\Delta_{QW}|$, or $\tau_B = \tau_{QW}$, we have $\delta_0 = 0$ and Eq. (2) is retrieved. The calculations give an inversion of the minibands. In order to discriminate the topological phase from the trivial one, the exact symmetry of the bound states at $q_z = 0$ and $|q_z| = \pi/L$ have been numerically calculated. The $L_6^+$ symmetry of the $j$-th confined states is calculated as $\int F_1^{(j)} F_1^{(j)} dz$, where the integral extends over one SL period: $0 \leq z \leq L$. One can then deduce the $L_6^-$ parity by $1 - \int F_1^{(j)} F_1^{(j)} dz$. A symmetry inversion is found when the state with a $L_6^+$-major component lies above the $L_6^-$-major state, and one can retrieve the Burkov-Balents phase diagram [3].

The results of the symmetry calculations are given in Fig. 2(b,c) of the main text. We deduce that at 4.2 K, both SL9-3.5 and SL27-1.5 display symmetry inverted miniband structure. This inversion is then experimentally demonstrated for SL27-1.5 (see Fig. 4). Oppositely, SL15-3.5 presents a normal symmetry order mainly because for its given layer thicknesses, $2\Delta_{QW} = -20$ meV is not enough.

## APPENDIX D: LANDAU LEVELS OF THE MINIBANDS

The $B$-dependent terms in (A1) are taken into account in a perturbation theory. The perturbative Hamiltonian for $n > 0$ is then:

$$\begin{pmatrix} 0 & 0 & 0 & v_\parallel\sqrt{2e\hbar Bn} \\ 0 & 0 & v_\parallel\sqrt{2e\hbar Bn} & 0 \\ 0 & v_\parallel\sqrt{2e\hbar Bn} & 0 & 0 \\ v_\parallel\sqrt{2e\hbar Bn} & 0 & 0 & 0 \end{pmatrix}$$

We derive an effective Hamiltonian expressed in the basis of the normalized envelope functions of $TMB$ and $TMB'$ obtained above at $k_x = k_y = 0$ and for a given $q_z$. It gives:

$$H^{eff}_{TMB-TMB'}(q_z) = \begin{pmatrix} -\delta(q_z) & 0 & 0 & Av_\parallel\sqrt{2e\hbar Bn} \\ 0 & -\delta(q_z) & Av_\parallel\sqrt{2e\hbar Bn} & 0 \\ 0 & Av_\parallel\sqrt{2e\hbar Bn} & \delta(q_z) & 0 \\ Av_\parallel\sqrt{2e\hbar Bn} & 0 & 0 & \delta(q_z) \end{pmatrix}$$

where $A = \int \left[ F_1^{(TMB)} F_2^{(TMB')} + F_1^{(TMB')} F_2^{(TMB)} \right] dz = \pm i$ by parity. Therefore, the twofold degenerated Landau levels for $n > 0$ are those of Dirac fermions:

$$E_n = \pm \sqrt{\delta^2(q_z) + 2e\hbar v_\parallel^2 Bn} \tag{E1}$$

In this system, the $n = 0$ Landau levels are spin-polarized and non-dispersive in magnetic field. The corresponding Landau levels are given in Fig. 6. In the present work, experiments have been performed in the Faraday geometry that leads to conventional dipole selection rules. Magneto-optical transitions are thus occurring between two Landau levels $n \to n \pm 1$ and at fixed $q_z$. For instance, the ground transition observed in Fig. 4 involves the levels 0 of $TMB'$ and 1 from $TMB$.

More subtly, we want to point out that the perturbative Hamiltonian is exact for the longitudinal valley whose high symmetry axis is naturally aligned with $B//[111]$; however, it is not the case for the oblique valleys, which main axis is tilted from $z$ by an angle $\theta = 70.5°$. This anisotropy effect has been considered by rotating spin and momentum operators in the Hamiltonian, an operation which is detailed in ref. [61]. This result allows us to adopt an empirical approach in this work by modeling the anisotropy effect with different valley-dependent fitting parameters $v_\parallel$ and $v_z$ in Eq (E1).

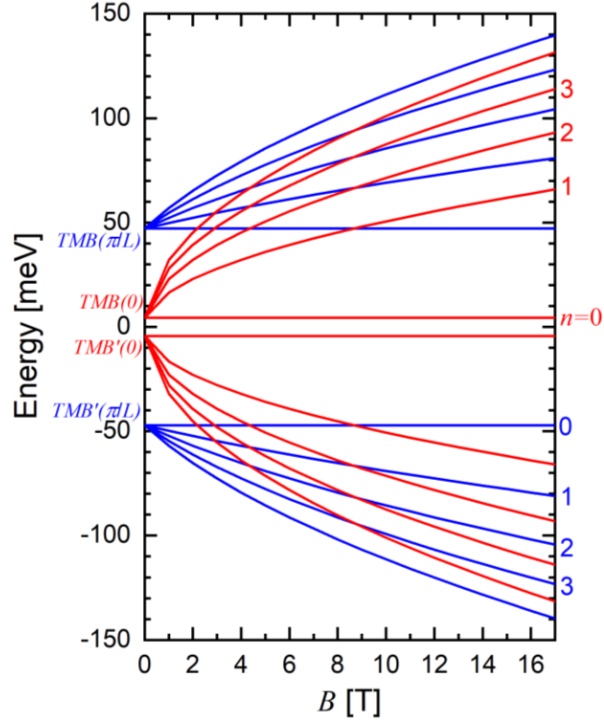

**Figure 6. Landau levels of the topological minibands.** Calculated Landau levels of $TMB$ and $TMB'$ at $q_z = 0$ (red) and $q_z = \pm\pi/L$ (blue). Calculations have been performed with the parameters $d_{QW} = 9$ nm; $d_B = 3.5$ nm; $2\Delta_{QW} = -72.5$ meV; $2\Delta_B = +150$ meV and $v_\parallel = v_z = 4.40 \times 10^5$ m/s. Some Landau level indexes $n$ are written at the right.

# Supplementary material for: Miniband engineering and topological phase transitions in topological - normal insulator superlattices

G. Krizman[1,2], B.A. Assaf[3], G. Bauer[2], G. Springholz[2], L.A. de Vaulchier[1], Y. Guldner[1]

[1] Laboratoire de Physique de l'Ecole normale supérieure, ENS, Université PSL, CNRS, Sorbonne Université, 24 rue Lhomond 75005 Paris, France

[2] Institut für Halbleiter und Festkörperphysik, Johannes Kepler Universität, Altenberger Strasse 69, 4040 Linz, Austria

[3] Department of Physics, University of Notre Dame, Notre Dame, IN 46556, USA

## Parameters determined from magneto-optical experiments

**Table S1.** Fitting parameters for the three SL samples at each temperature. N.A is for "Not accessible". The error bars are the following: $\pm 2.5$ meV for $2\Delta_{QW}$ ; $\pm 0.05 \times 10^5$ m/s for $v_\parallel^l$ ; $\pm 0.05 \times 10^5$ m/s for $v_\parallel^o$ ; $\pm 0.10 \times 10^5$ m/s for $v_z^l$ ; $\pm 0.10 \times 10^5$ m/s for $v_z^o$. The error bars for $2\Delta_B$ are drawn on Fig. 5(a,b,c) of the main text.

| Samples | $T$ [K] | $2\Delta_{QW}$ [meV] | $2\Delta_B$ [meV] | $v_\parallel^l$ [m/s] | $v_z^l$ [m/s] | $v_\parallel^o$ [m/s] | $v_z^o$ [m/s] |
|---|---|---|---|---|---|---|---|
| SL9-3.5  | 4.2 | -72.5 | 150 | 4.40 | 4.40 | 4.10 | 4.70 |
|          | 80  | -55   | 165 | 4.40 | 4.40 | 4.10 | 4.70 |
|          | 120 | -40   | 190 | 4.40 | 4.40 | 4.10 | 4.70 |
|          | 160 | -12.5 | 225 | 4.45 | 4.45 | 4.10 | 4.70 |
|          | 200 | 10    | 255 | 4.50 | 4.50 | 4.10 | 4.70 |
| SL15-3.5 | 4.2 | -20   | 150 | 4.50 | 4.50 | 4.10 | 4.70 |
|          | 50  | -10   | 155 | 4.50 | 4.50 | 4.10 | 4.70 |
|          | 80  | -2.5  | 165 | 4.50 | 4.50 | 4.10 | 4.70 |
|          | 120 | 15    | 180 | 4.50 | 4.50 | 4.10 | 4.70 |
|          | 160 | 32.5  | 200 | 4.50 | 4.50 | 4.10 | 4.70 |
|          | 200 | 55    | 220 | 4.50 | 4.50 | N.A  | N.A  |
| SL27-1.5 | 4.2 | -60   | 140 | 4.45 | 4.45 | 4.45 | 4.45 |
|          | 20  | -57.5 | 145 | 4.45 | 4.45 | 4.45 | 4.45 |
|          | 40  | -50   | 150 | 4.45 | 4.45 | 4.45 | 4.45 |
|          | 80  | -32.5 | 155 | 4.40 | 4.40 | 4.40 | 4.40 |
|          | 120 | -15   | 175 | 4.35 | 4.35 | 4.35 | 4.35 |
|          | 160 | 0     | 200 | 4.35 | 4.35 | 4.35 | 4.35 |
|          | 200 | 10    | 230 | 4.35 | 4.35 | 4.35 | 4.35 |

## Miniband dispersions of SL9-3.5 and SL15-3.5 at T=4.2 K

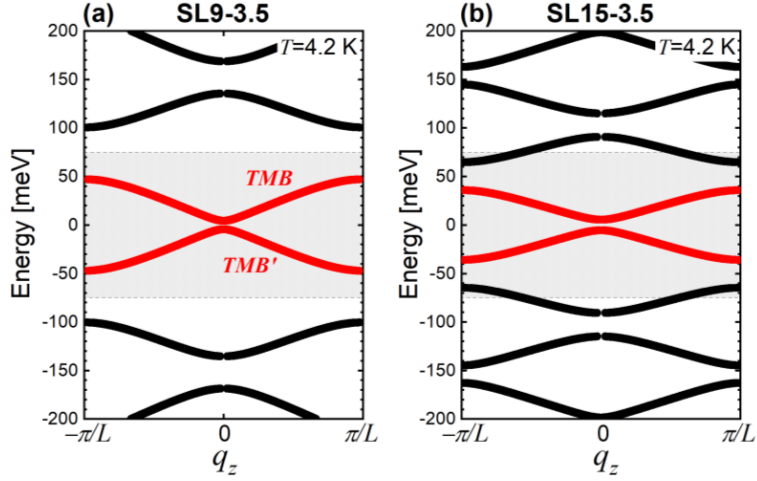

**Figure S1.** Miniband dispersions of SL9-3.5 and SL15-3.5 at T=4.2 K. In red are the calculated $TMB$ and $TMB'$ and in black are the higher excited minibands. The grey area indicates the barrier potential height.

## Miniband inversion in the vicinity of the phase transition

In a simplified picture of a 3D Dirac material, the trivial ordering is restored when $\hbar v_D k \sim 2|\Delta|$, thus, occurring at $k \sim 0.025$ Å$^{-1}$ for $2|\Delta| \sim 60$ meV. In the investigated SLs, the wave vector of the mini BZ is limited to π/L which does not exceed 0.025 Å$^{-1}$ even for the SL9-3.5 with the smallest SL period. Thus, the condition of full miniband band inversion is fullfilled for our samples SL9-3.5 and SL27-1.5 at low temperature $T < T_C$. At higher temperatures, however, there exists a tiny region of partially inverted minibands near the topological phase transition where the miniband gap gets very small, and thus, the above-mentioned limiting condition is met before the zone edges. This is illustrated in the Fig. S2 below, where we show the partially inverted minibands of the sample SL9-3.5 at ~40 K where the band gap is only a few meV.

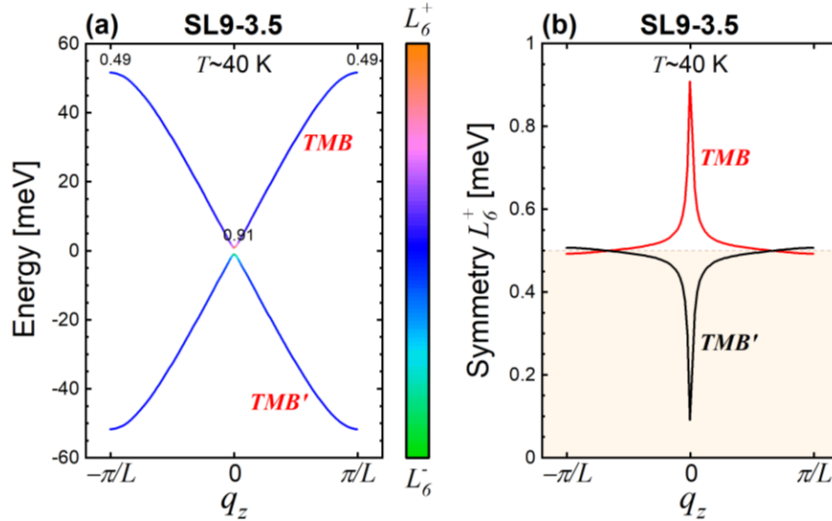

**Figure S2. (a)** Miniband dispersion for SL9-3.5 similar as Fig. 2(b) of the main text, but for $T \sim 40$ K, where the SL gap is negative but nearly zero, illustrating the case where the band inversion does not occur over the whole mini BZ but only in a very narrow region around $q_z$ =0 (see color scale and labels which denote the $L_6^+$-parity). **(b)** Calculated $L_6^+$-parity of TMB and TMB' shown in (a). The miniband is only partially inverted around $q_z$ =0.

21